\documentclass[%
 reprint,
superscriptaddress,
amsmath,amssymb,
aps,
prb,
]{revtex4-2}

\usepackage{graphicx}%
\usepackage{multirow}%
\usepackage{amsmath,amssymb,amsfonts}%
\usepackage[normalem]{ulem}
\usepackage{amsthm}%
\usepackage{mathrsfs}%
\usepackage[title]{appendix}%
\usepackage{xcolor}%
\usepackage{textcomp}%
\usepackage{booktabs}%
\usepackage{algorithm}%
\usepackage{algorithmicx}%
\usepackage{listings}%
\usepackage{comment}

\usepackage{graphicx}
\usepackage{dcolumn}
\usepackage{bm}

\begin{document}

\title{Coherent and dissipative coupling in a magneto-mechanical system}

\author{P. Carrara}
\thanks{These two authors contributed equally}
\affiliation{Dipartimento di Fisica, Università degli Studi di Milano, Via Celoria 16, 20133 Milano, Italy}
\affiliation{Cnr-Istituto Officina dei Materiali, Unità di Trieste, Strada Statale 14, km 163.5, 34149 Basovizza (TS), Italy}

\author{M. Brioschi}
\thanks{These two authors contributed equally}
\affiliation{Dipartimento di Fisica, Università degli Studi di Milano, Via Celoria 16, 20133 Milano, Italy}
\affiliation{Cnr-Istituto Officina dei Materiali, Unità di Trieste, Strada Statale 14, km 163.5, 34149 Basovizza (TS), Italy}

\author{R. Silvani}
\affiliation{Dipartimento di Fisica e Geologia, Università di Perugia, Via A. Pascoli, 06123 Perugia, Italy}

\author{A.O. Adeyeye}
\affiliation{Department of Physics, Durham University, South Rd, DH1 3LE Durham, United Kingdom}

\author{G. Panaccione}
\affiliation{Cnr-Istituto Officina dei Materiali, Unità di Trieste, Strada Statale 14, km 163.5, 34149 Basovizza (TS), Italy}

\author{G. Gubbiotti}
\email{Corresponding author, email: gubbiotti@iom.cnr.it}
\affiliation{Cnr-Istituto Officina dei Materiali, Unità di Perugia, Via A. Pascoli, 06123 Perugia, Italy}

\author{G. Rossi}
\email{Corresponding author, email: giorgio.rossi2@unimi.it}
\affiliation{Dipartimento di Fisica, Università degli Studi di Milano, Via Celoria 16, 20133 Milano, Italy}
\affiliation{Cnr-Istituto Officina dei Materiali, Unità di Trieste, Strada Statale 14, km 163.5, 34149 Basovizza (TS), Italy}

\author{R. Cucini}
\affiliation{Cnr-Istituto Officina dei Materiali, Unità di Trieste, Strada Statale 14, km 163.5, 34149 Basovizza (TS), Italy}

\begin{abstract} 
Hybrid elastic and spin waves hold promises for energy-efficient and versatile generation and detection of magnetic signals, with potentially long coherence times.
Here we report on the combined elastic and magnetic dynamics in a one-dimensional magneto-mechanical crystal composed of an array of magnetic nanostripes.
Phononic and magnonic modes are impulsively excited by an optical ultrafast trigger and their decay is monitored by time resolved Magneto Optical Kerr Effect.
Complementary Brillouin Light Scattering measurements and micromagnetic simulations concur in a unified picture in which the strength and degree of mixing of coherent and dissipative coupling of the quasi-particles are determined quantitatively. 
\end{abstract}

\maketitle

In hybrid magnonics, the coupling of magnons (the quanta of spin waves) to other degrees of freedom is explored to achieve enhanced functionalities in solid-state systems and devices \cite{lachance2019hybrid}.
Novel results have been achieved in coupling magnons to microwave \cite{zhang2014strongly} and optical photons \cite{liu2016optomagnonics}, and/or to phonons \cite{zhang2016cavity,berk2019strongly,li2021advances, carrara2022all}, or to superconducting qubits \cite{lachance2017resolving}.
This approach has found fertile ground in the field of quantum engineering \cite{clerk2020hybrid}, where hybridized quasi-particles can boost transduction and sensing capabilities \cite{verhagen2012quantum,nair2021enhanced}, down to the single-quantum detection and manipulation \cite{tabuchi2014hybridizing}, or allow novel computation, simulation and storage platforms \cite{chumak2022advances}.

Here we report the experimental observation of magnon-phonon hybridization in a 1D magnonic-phononic crystal via time-resolved Magneto-Optical Kerr Effect (tr-MOKE), namely the building of a Von Neumann$-$Wigner hybridization gap \cite{von1929uber} in a quasiparticle-quasiparticle system in the solid state.
We obtain experimental evidence of the mixture of coherent and dissipative coupling, a condition we dub here \textit{mixed} coupling. A Hamiltonian model for the hybridized modes corroborates our findings.
The coherent and dissipative coupling \cite{harder2021coherent} describes the energy exchange between two coherently coupled systems directly \cite{hioki2022coherent}, or via correlated dissipation into a common reservoir \cite{harder2018level}, respectively. Dissipative coupling is now boosting research for its close link to non-Hermitian physics \cite{yang2020unconventional, li2023exceptional} and non-reciprocal transport \cite{wang2019nonreciprocity}.

Both coupling mechanisms can coexist in a single system, creating the possibility of tuning one into the other.
A few papers report mixed coupling in magnon-photon hybridization \cite{wang2014interplay,harder2018level,zhang2018coherent,wang2019nonreciprocity}; here we expand on this by reporting for the first time mixed coupling in magnon-phonon hybrid.
Experiments in time-domain prove crucial to explore such physics in the weak-coupling regime.

\begin{figure*}
\centering
\includegraphics[width =.8\textwidth]{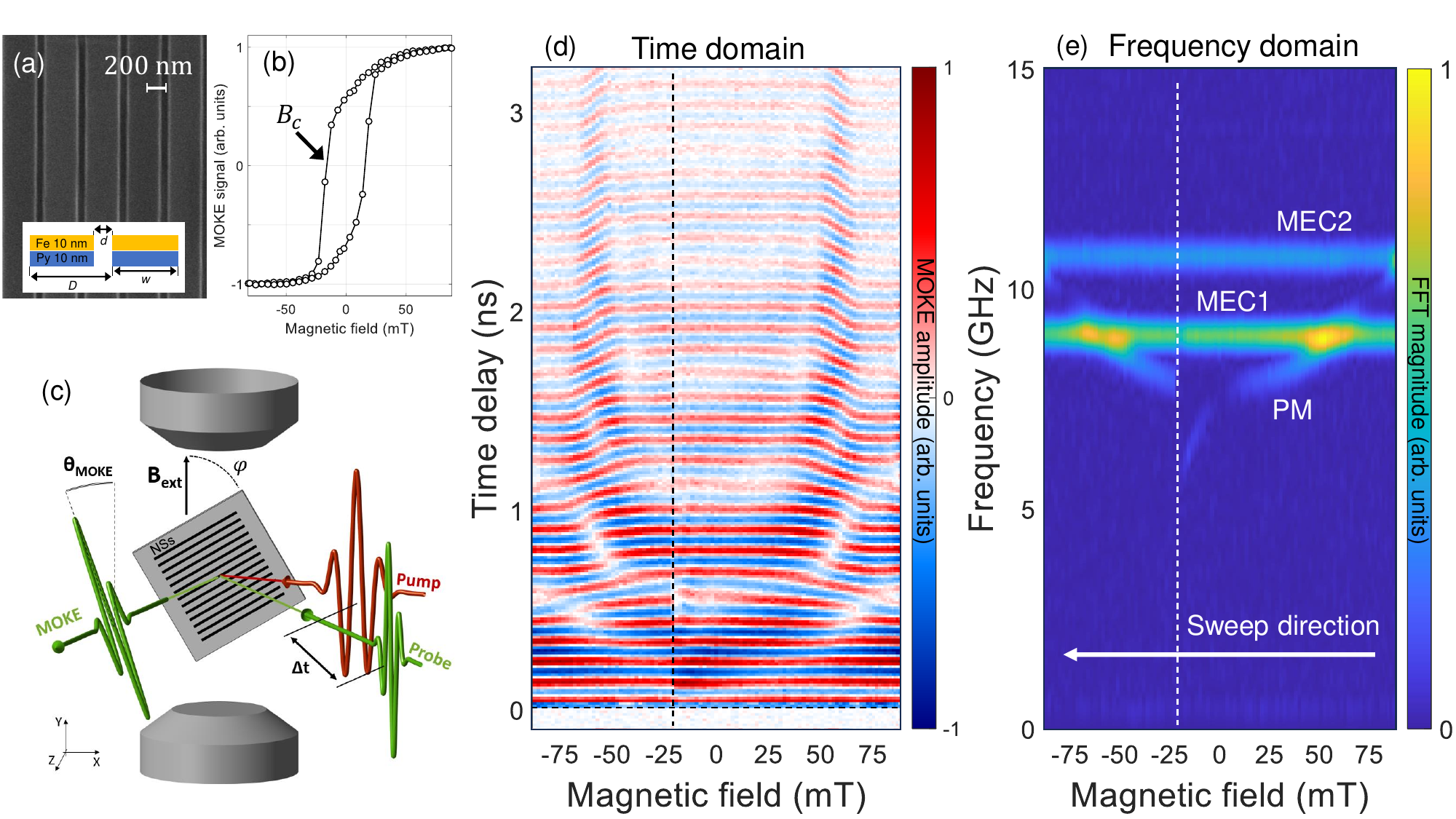}
\caption{(a) SEM micrograph of the investigated sample. Inset: a cross-sectional sketch of the bilayered NSs; the inter-NS distance $d$, NS width $w$, and full periodicity $D$ are also indicated (not to scale). (b) Magnetic hysteresis loop obtained via static MOKE at $\varphi = 60^{\circ}$; the coercive field $B_c = 17$ mT is indicated. (c) Sketch of the experimental setup for tr-MOKE. (d-e) Time- and frequency-domain maps of the tr-MOKE signal at $\varphi = 60^{\circ}$. The magnetic field is swept from positive to negative values. The vertical dashed lines at $B_\text{ext} = -22$ mT highlight the magnetization reversal, which results in phase anomalies in the time domain and in discontinuity for the PM mode in the frequency domain. The enhancement of MOKE amplitude at late delays in panel (d) corresponds to the condition of PM-MEC1 coupling in panel (e).}
\label{Fig:General}
\end{figure*}

The sample employed in this study is an array of rectangular cross-sectioned bilayered nanostripes (NSs) of Fe (10 nm thick) and Py (Ni$_{80}$Fe$_{20}$, 10 nm thick) on Si (001) substrate. Details on the fabrication process can be found in Ref.\cite{adeyeye2008large}.
Each NS is $w = 340$ nm wide, and the inter-NS spacing is $d = 70$ nm; this gives an overall periodicity $D = w+d = 410$ nm (Fig.\ref{Fig:General}(a)). 
The coercive field is $B_c = 17$ mT, as extracted from static MOKE hysteresis loops (Fig.\ref{Fig:General}(b)).
Such system has been characterized as a magnonic crystal via Brillouin Light Scattering (BLS), ferromagnetic resonance and micromagnetic simulations \cite{gubbiotti2016collective,kostylev2016microwave,silvani2018spin,demand2002}: magnon bands form as for inter-NS magnetic dipolar interaction.
Similarly, the spatial modulation of the elastic properties of the sample surface gives rise to standing surface phononic modes, together with acoustic modes localized in each NS \cite{maznev2011mapping,pan2013phononic,ma2023phonon}.

The optical end-station employed for pump-probe spectroscopy is sketched in Fig.\ref{Fig:General}(c): the sample is excited with an ultrafast near-infrared pulsed laser (pump), and the magnetic and magneto-elastic dynamics are probed via tr-MOKE.
The incidence angles from the sample normal are $12^\circ$ and $6^\circ$ for the pump and probe beams, respectively; consequently, we are primarily sensitive to the out-of-plane component of the magnetization.
Further details on the setup are reported in Ref.\cite{brioschi2023multidetection} and in the Supplemental Material \cite{SupplementaryInformation}.
The pump excitation triggers acoustic modes and, via inverse magnetostriction, the phononic strain fields interact with the magnetization of the NSs, leading to  coupled magneto-mechanical dynamics \cite{godejohann2020magnon}.
Moreover, it is possible to excite pure magnetization dynamics via ultrafast heat-induced magnetic anisotropy quenching upon ultrafast laser illumination.
This thermal mechanism, firstly described in Ref.\cite{van2002all} for an easy-plane magnetic anisotropy system and then extended to different systems \cite{chang2018selective,blank2022laser}, requires an external magnetic field with strength comparable to the sample magnetic anisotropy field. 
The magnon modes excited with this mechanism rely only on the magnetostatic properties of the NSs (magnetization and shape anisotropy) and can be observed via complementary techniques like BLS, as well as simulated with micromagnetic models.

In Fig.\ref{Fig:General}(d) we report a wide-scan map of tr-MOKE results, for $B_\text{ext}$ swept from +90 to -90 mT, in approximately 1.5 mT steps.
For each magnetic field, we record the tr-MOKE signal up to a delay of 3.3 ns.
An exponential plus linear background is subtracted from each trace.
The map in Fig.\ref{Fig:General}(e) is obtained performing FFT of each trace. 
Here, three features can be identified: two flat modes with different intensities, featuring no frequency dispersion with $B_\text{ext}$, and a third dispersive mode.
We label the formers as magneto-elastic-coupling modes (MEC1 and MEC2) and the latter as a pure magnonic mode (PM).
We now briefly discuss their excitation mechanism.

The pump photons are absorbed by the metallic NSs, resulting in thermo-elastic expansion (the direct bandgap of Si is well above the pump photon energy).
The periodic strain field stabilizes a standing surface acoustic wave with the wavelength matching the NS array periodicity ($D$).
Pump-induced heating also generates localized breathing modes within each NS.
These acoustic modes drive NS magnetization, producing the time-modulated magnetic contrast we observe in tr-MOKE (flat modes in Fig.\ref{Fig:General}(e)).

\begin{figure}
\centering
\includegraphics[width = 0.45\textwidth]{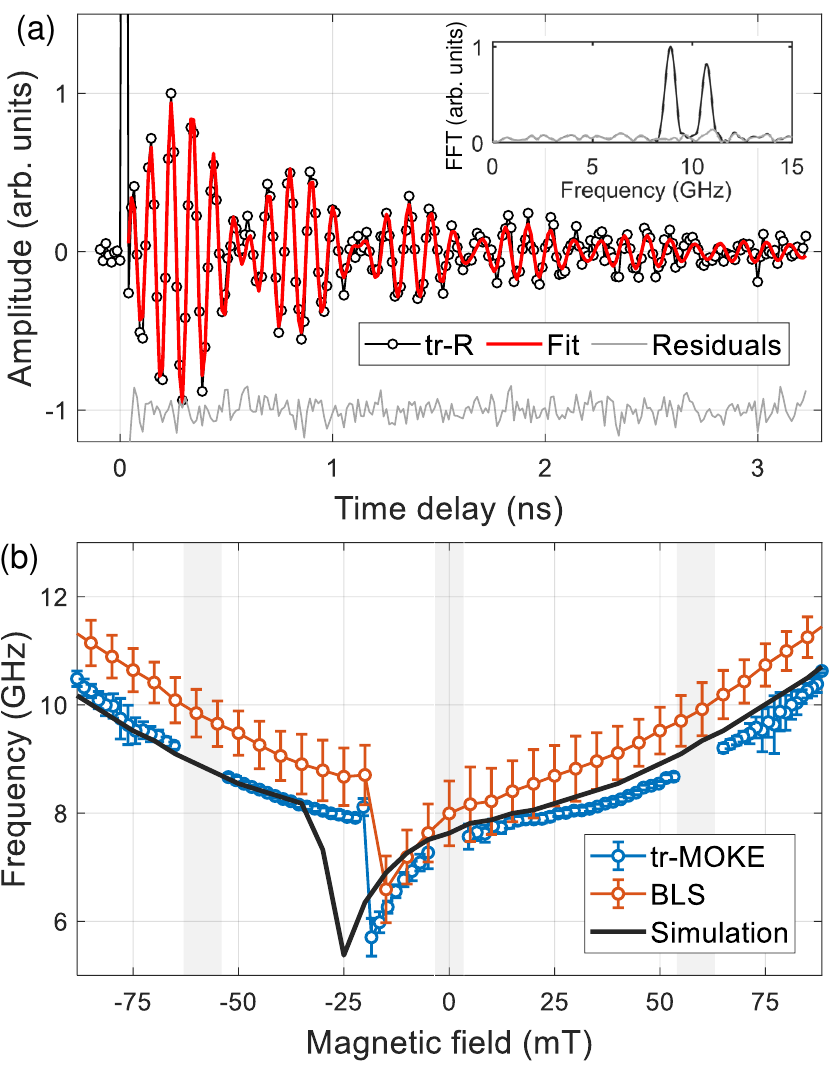}
\caption{(a) Time-domain trace (black circles) and fit (red line) of time-resolved reflectivity at $B_\text{ext} = 0$ mT. The residuals (grey line) are rigidly shifted for clarity. Inset: FFT of the original trace (black line) and of the residuals (grey line). (b) The frequency of the PM mode was extracted from a time-domain fit of tr-MOKE traces (blue circles), together with the frequency of the lowest-order magnon mode as obtained from BLS measurements (orange circles) and micromagnetic simulations (black line). The azimuth is $\varphi = 60^\circ$. The three vertical shaded bars indicate regions in which the extraction of frequency from tr-MOKE data is not possible.}
\label{Fig:Attribution}
\end{figure}

The attribution of such flat modes to magneto-elastic modes in similar systems is consistent with the literature \cite{chang2018selective,godejohann2020magnon}.
To further confirm the assignment, we perform time-resolved reflectivity measurements (tr-R) using identical experimental conditions as in the tr-MOKE measurements, while detecting the non-rotated component of the probe polarization.
The time-domain trace after background removal is reported in Fig.\ref{Fig:Attribution}(a) (black circles), for $B_\text{ext} = 0$ mT.
The fit of the function
\begin{equation}
y = \sum_i A_i \sin{(\omega_i t + \phi_i)}e^{-\gamma_i t}
\label{Eq:Osc}
\end{equation}
is shown in red ($i = 1,2$ denotes the lower- and upper-frequency phononic mode, respectively).
The parameters $A_i,~\omega_i,~\phi_i,~\gamma_i$ are the amplitude, angular frequency, phase and damping parameter for $i$-th mode.
The feature-less residuals of the fitting are shown in grey and rigidly shifted for clarity.
The frequencies ($f_i = \omega_i/2\pi$) and dampings obtained as best-fit parameters are $f_\text{MEC1} = (8.89 \pm 0.05)$ GHz and $f_\text{MEC2} = (10.75 \pm 0.05)$ GHz, and $\gamma_\text{MEC1} = (0.73 \pm 0.09)$ rad/ns and $\gamma_\text{MEC2} = (1.4 \pm 0.1)$ rad/ns, in agreement with the parameters for the flat modes in tr-MOKE (see below).
The lower frequency mode is assigned to the Rayleigh wave of the substrate covered with the NS array \cite{chang2018selective}, and the higher frequency mode to a localized (width) breathing mode of each NS.

Let us now focus on the PM mode.
The leading magnetic anisotropy in the NSs is shape anisotropy, which for thin NSs favors in-plane magnetization along the NS longitudinal axis, which is thus the magnetic easy axis (EA).
If $B_\text{ext}$ is not aligned to EA and comparable in strength to the magnetic anisotropy field, the pump absorption in the NSs results in ultrafast quenching of the magnetization and concurrently to impulsive softening of the magnetic anisotropy, triggering magnetization precession; dynamics at the GHz range reflects the equilibrium magnon eigenstates of the system \cite{SupplementaryInformation,van2002all}.
Furthermore, the same triggered dynamics in each NS result in synchronized precession, generating a measurable MOKE signal—a zero-wavevector magnonic mode.
Note that the PM mode spectral weight diminishes to zero whenever the magnetization equilibrium axis aligns with EA: (\textit{i}) as $B_\text{ext}$ weakens (see Fig.\ref{Fig:General}(e)), or (\textit{ii}) at $\varphi = 0^{\circ}$ for any $B_\text{ext}$ value \cite{SupplementaryInformation}.
To confirm the PM mode assignment as the lowest-order magnetization precession, we compare tr-MOKE with BLS and micromagnetic simulations (see \cite{SupplementaryInformation} for details). In Fig.\ref{Fig:Attribution}(b), we present the PM mode frequency from tr-MOKE fits (blue circles) alongside the lowest-order magnonic frequency's field dependence from BLS (orange circles) and micromagnetic simulations (black line). The datasets and simulations show good agreement, especially in the concavity change at positive field and switching field value. The grey shaded bars in Fig.\ref{Fig:Attribution}(b) highlight regions in which extraction of tr-MOKE values is not possible, either because of the absence of the PM mode or because of the mixing to the MEC modes. The systematic redshift of the tr-MOKE results is compatible with a few degrees of experimental mismatch in the azimuth angle $\varphi$.

\begin{figure}
\centering
\includegraphics[width = 0.45\textwidth]{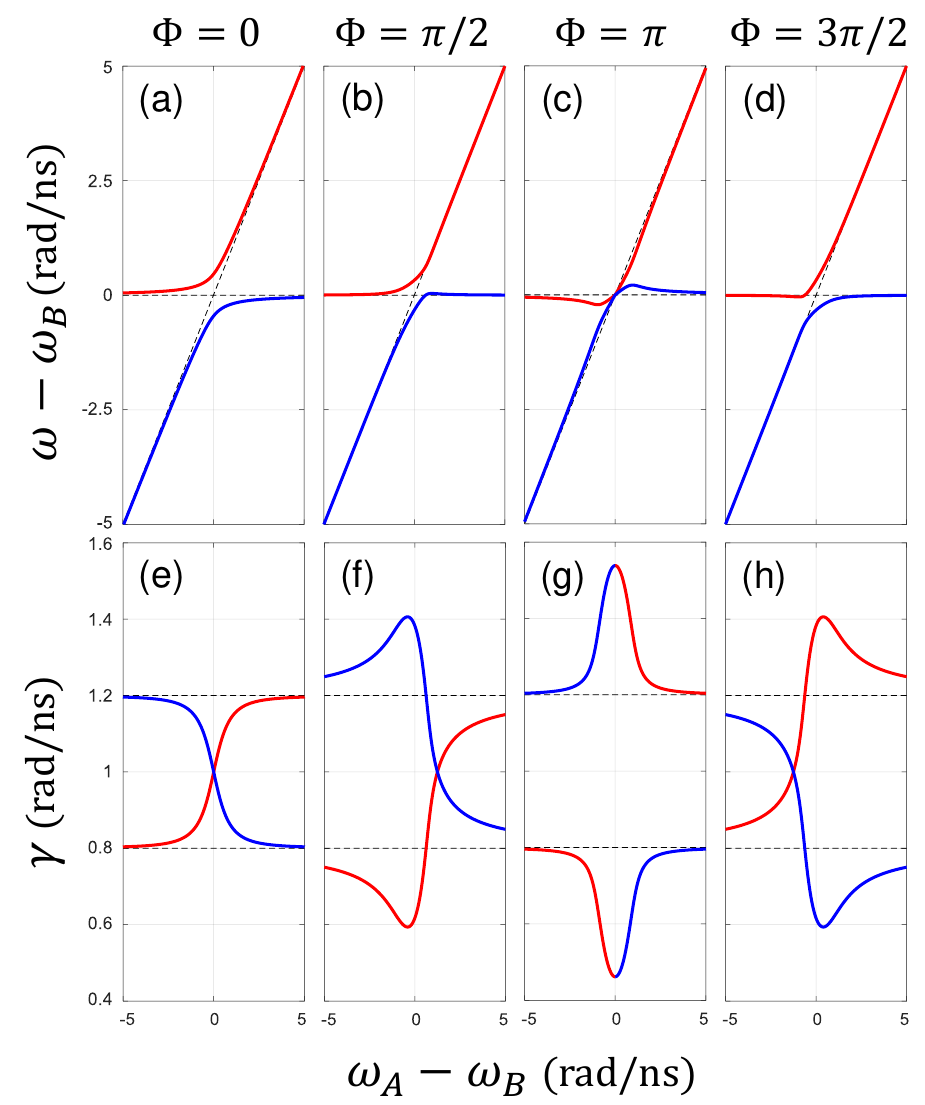}
\caption{Modification of the eigenvalues of $\mathcal{H}$ upon changing the coupling phase $\Phi$.
(a-d) Frequency dispersions (real part).
(e-h) Damping parameters (imaginary part).
All plots are shown as a function of the uncoupled frequency detuning.
The upper ($\tilde{\omega}_+$) and the lower branch ($\tilde{\omega}_-$) results are shown as red and blue lines, respectively.
The uncoupled values for frequency and damping are shown in each plot as black dashed lines.
For comparison to the experimental results shown later, in the calculations we assumed $\gamma_A = 1.2$ rad/ns and $\gamma_B = 0.8$ rad/ns; the coupling strength was kept at $g = 0.5$ rad/ns.}
\label{Fig:eigenvalues}
\end{figure}

We now focus on the region (50 - 80) mT, where the crossing of the PM and MEC1 modes (see Fig.\ref{Fig:General}(e)) suggests the presence of a coupling of magneto-elastic origin.
To gain a deeper insight, we employ a Hamiltonian $\mathcal{H}$, modeling both coherent and dissipative coupling \cite{harder2018level,xu2019cavity}:
\begin{equation}
    \mathcal{H} / \hbar = \tilde{\omega}_A a^\dag a + \tilde{\omega}_B b^\dag b + g \left( a^\dag b + e^{i\Phi} a b^\dag \right)~.
    \label{Eq:hamiltonian}
\end{equation}
Here $a^\dag$ and $b^\dag$ ($a$ and $b$) are the creation (annihilation) operators for mode $A$ and $B$, respectively (in our experiment the modes are the phonon and the magnon); $\tilde{\omega}_i = \omega_i - i\gamma_i$ is the generalized angular frequency of
uncoupled mode $i = (A, B)$, encompassing both the angular frequency $\omega_i$ and the intrinsic damping $\gamma_i$; $g$ is the strength of the coupling, whose nature depends on the value of the phase $\Phi$.
The coupled eigenvalues of $\mathcal{H}$ are
\begin{equation}
    \tilde{\omega}_\pm = \left( \frac{\tilde{\omega}_A + \tilde{\omega}_B}{2} \right) \pm \sqrt{\left( \frac{\tilde{\omega}_A-\tilde{\omega}_B}{2} \right)^2 + g^2 e^{i\Phi}}~,
    \label{Eq:eigenvalues}
\end{equation}
where again the real part of $\tilde{\omega}_\pm$ gives the angular frequency and the opposite imaginary part gives the damping parameter.
In Fig.\ref{Fig:eigenvalues} the coupled (red and blue lines) and uncoupled (black dashed lines) eigenvalues are plotted as a function of the uncoupled frequency detuning.
At $\Phi = 0$ (pure coherent coupling, panels (a) and (e)), frequency gapping at zero detuning results; this is often understood as the \textit{smoking gun} for hybridization.
The dampings, on the other hand, attract each other and are equal at zero detuning.
At $\Phi = \pi$ (pure dissipative coupling, panels (c) and (g)) the deviation from the uncoupled values are opposite: frequency attraction extends the degeneracy to a finite region across zero detuning, while the dampings repel each other.
The eigenvalue symmetry for $\Phi = 0$ and $\pi$ is lost if both couplings are at play, \textit{i.e.} for mixed coupling, as shown for $\Phi = \pi/2$ (panels (b) and (f)) and $\Phi = 3\pi/2$ (panels (d) and (h)).
The striking feature in mixed coupling is that the condition for degenerate dampings is shifted from zero detuning.
We propose this shift as a phenomenological identifier of mixed coupling systems.

\begin{figure}
\centering
\includegraphics[width = 0.4\textwidth]{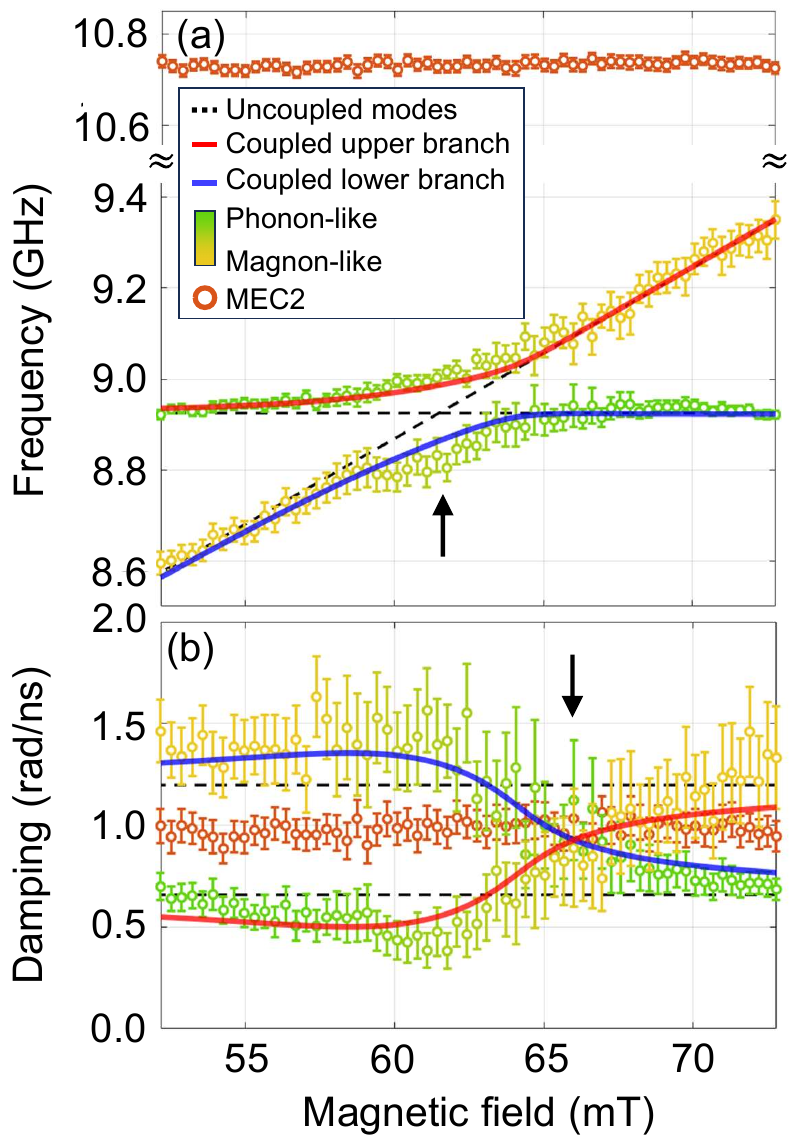}
\caption{Analysis of frequency (a) and damping (b) as extracted from the time-domain fit of tr-MOKE traces acquired in a close-up of the PM-MEC1 crossing region. The experimental data for the hybridizing modes show deviations from the uncoupled modes, mixing phonon character (field-independent frequency, green circles) and magnon character (field-dependent frequency, yellow circles). The fit results for the upper hybridized branch (red line) and for the lower hybridized branch (blue line) are also shown, together with the values for the uncoupled modes (black dashed lines). The arrows highlight the mismatch between the field values for zero-detuning (upper panel) and for damping crossing (lower panel). The fit results for the non-hybridizing MEC2 mode are also reported (orange circles).}
\label{Fig:crossingFit}
\end{figure}

The proposed Hamiltonian model requires the time domain fit of the traces, acquired in finely sampled intervals (approximately 0.3 mT) of $B_\text{ext}$ in a close-up of the PM-MEC1 crossing region.
In Fig.\ref{Fig:crossingFit} we report, as color-coded circles with error bars, the frequency (panel (a)) and damping (panel (b)) obtained as best-fit parameters. Note that the time-domain analysis reveals frequency gapping of the modes, a feature not visible in the FFT map (see also \cite{SupplementaryInformation}).

As highlighted by the black arrows, there is a mismatch between the damping crossing and the condition of zero detuning.
We fit to the data in Fig.\ref{Fig:crossingFit} the eigenvalues of $\mathcal{H}$ (Eq.\ref{Eq:eigenvalues}): the best-fit curves are reported for the upper branch (red line) and for the lower branch (blue line).
The PM mode is assumed as linear in $B_\text{ext}$, a reasonable approximation of the actual non-linear dispersion, given the reduced detuning window.
The result of the fitting gives $g = \left( 0.55 \pm 0.03 \right)$ rad/ns and $\Phi = 1.3 \pm 0.1$. 
These values set the investigated system as weakly coupled since the strength $g$ is lower than the intrinsic damping of both hybridizing modes; as $\Phi$ is close to $\pi/2$, both coupling mechanisms are at play with comparable strength.

To summarize, we derived from tr-MOKE experimental data the coupling of phononic and magnonic modes in a 1D magnonic-phononic crystal.
The time-domain data give evidence of a mixed phonon-magnon coupling mechanism at play.
By means of comparison of the experimental data with the eigenvalues of a comprehensive Hamiltonian model we derived the coupling strength and the competition between coherent and dissipative coupling.
Such quantitative information will be crucial for identifying the experimental parameters that continuously tune the coupling in a magneto-mechanical system, transitioning from purely coherent to purely dissipative coupling.
Our results suggest magnonic-phononic crystals as ideal platforms to investigate magnon-phonon mixed coupling, and hint to the possibility of novel magnonic-phononic devices, in analogy to what was shown by cavity magnonics (magnon-photon hybridization) \cite{harder2018level,xu2019cavity}.
Moreover, we propose the observed frequency-damping mismatch as an experimental identifier for such mixed coupling: this is not limited to magnon-phonon coupling, rather can be a useful discriminator also for other hybridizing platforms.
Finally, we demonstrated time-domain spectroscopy as a key tool to investigate weakly coupled hybrid systems, allowing for reliable quantitative analysis of the coupling even when intrinsic damping dominates over coupling strength, challenging frequency-domain techniques.

P.C. and M.B. thank Vincent Polewczyk and Giacomo Jark for valuable discussions.
P.C. is also grateful to Riccardo Panza and Alberto Scazzola for insightful discussions.
This work is performed in the framework of the Nanoscience Foundry and Fine Analysis (NFFA-MUR Italy Progetti Internazionali) facility.
Research at IOM-CNR has been also funded by the European Union - Next Generation EU under the Italian Ministry of University and Research (MUR) National Innovation Ecosystem grant ECS00000041 - VITALITY-CUP B43C22000470005.
G.G. and G.P. acknowledge Università degli Studi di Perugia, CNR and MUR for support within the project Vitality.
A.O.A. and G.G. acknowledge the funding from the Royal Society through the Wolfson Fellowship and International Exchanges IEC\textbackslash R2\textbackslash 222074.

%

\end{document}